\documentclass[aps,prl,twocolumn,groupedaddress,floatfix,amsmath,amssymb]{revtex4}

\usepackage{graphicx}
\usepackage{bm}
\begin{document}

\title{Electronic and Quantum Transport Properties of Substitutionally Doped Double-Walled Carbon Nanotubes}

\author{Alejandro Lopez-Bezanilla$^{1}$}
\email[]{alejandrolb@gmail.com}
\affiliation{$^{1}$Materials Science Division, Argonne National Laboratory, 9700 S. Cass Avenue, Lemont, IL 60439, USA}

\begin{abstract}
A first-principles investigation of the electronic and quantum transport properties of double-walled carbon nanotubes doped with nitrogen and boron atoms is presented.
Concentric nanotube sidewalls separated by the typical graphitic van der Waals bond distance are found to strongly interact upon incorporation of doping atoms in the hexagonal networks.
The local perturbation caused by the doping atoms extends over both shells due to a hybridization of their electronic states, yielding a reduction of the backscattering efficiency as compared with two independent single-walled nanotubes. A multi-scale approach for the study of transport properties in micrometer-long double-walled carbon nanotubes allows to demonstrate that transitions from the ballistic to the localized regime can occur depending on the type of doping and the energy of the charge carrier. These results shed light on the quantum mechanism by which B and N doping represent an efficient method to create mobility gaps in metallic concentric carbon nanotubes.
\end{abstract}

\maketitle
The unrelenting rise of nanoscale strategies employing carbon-based materials for substituting modern semiconductor technology based on Si is obtaining great benefits \cite{4399979}.
Carbon nanotubes (CNTs) and graphene have led to applications in novel electronic devices, making carbon a promising candidate for the fundamental building block of a new generation of electronic devices \cite{ISI:000316154900014}. 
The delocalized network of C atom $\pi$-orbitals and the absence of efficient electron-phonon interactions are 
ultimately related to the capacity of these materials 
to transport high electric current density, thereby reducing the power that is required to operate ultrathin devices \cite{RevModPhys.79.677}.
However, as transistor-based electronics approach the atomic scale, small amounts of disorder begin to have dramatic negative effects.
The modification of nanotube sidewalls may result in the creation of scattering centers that degradate the CNT conductance, yielding the total or partial loss of its conducting properties. 
Avriller et al. \cite{PhysRevB.74.121406} showed that in disordered N-doped single-walled CNTs the electron mean free path can decrease down to the full suppression of the tubes conductance. 
Surprisingly, one of the most promising pathways out of this quandary may emerge from recent efforts to introduce defects to construct high-performance devices. Indeed, far from representing a drawback in the design of an electronic device, the creation of mobility gaps as a result of the electron backscattering with impurities has been proposed as an efficient method to overcome the absence of an energy band gap in metallic materials, which prevents any electronic current flow from switching off \cite{PSSB:PSSB201000135}.

Individual tubes in multi-walled CNTs are well insulated from each other so that the current flows mainly along individual shells, with a total current capability larger than that of a single-walled CNT\cite{PhysRevLett.93.176806}. The intershell interaction and incommensurability effects of concentric walls may alter the ballistic conduction of independent tubes \cite{ISI:000169703600016} although, as for single-walled CNTs, the major source of backscattering comes from structural defects (vacancies) or chemical modifications.  
Experimental investigations concluded that the electrical current in a double-walled CNT (DWNT) whose outer sidewall was functionalized is comparable to the current carried by a pristine single-walled CNT, namely the inner tube is electrically active upon outer sidewall modification \cite{doi:10.1021/nn201024u}. Although experimental studies have provided some insight on transport mechanism of chemically modified multi-walled CNTs \cite{PhysRevB.63.161404}, a systematic theoretical study of the transport properties of doped DWNTs is needed.

\begin{figure}[htp]
 \centering
 \includegraphics[width=0.45 \textwidth]{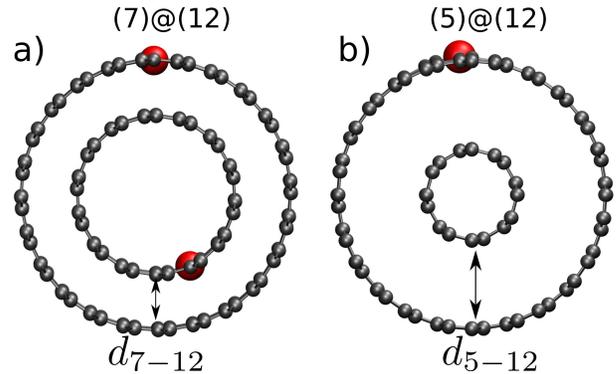}
 \caption{Two atomic structures of double-walled carbon nanotubes (DWNT) with substitutional doping. a) (7)@(12) DWNT with two foreign atoms in substitution of two C atoms at both the outer and inner-sidewalls. The inter-wall distance $d_{7-12}$ is $\sim$ 3.3 \AA. b) (5)@(12) DWNT with a doping atom in substitution of a C atom at the outer-sidewall tube. The inter-wall distance $d_{5-12}$ is $\sim$ 5.5 \AA .}
 \label{fig1}
\end{figure}

In this paper we investigate the distortion of the hyperconjugated network of two coaxial CNTs upon substitutional B and N doping, and the subsequent limitation on the conductance efficiency due to the creation of quasi-bound states. An analysis of the transmission as a function of the doping rate and the interwall separation between coaxial nanotubes is performed to study the sensitivity of relative shell distance on the quantum transport features. To this end, the parametrization formerly employed in pioneering studies \cite{adessi} of transport properties of single-walled CNTs is insufficient. Here we use a multi-scale approach based on first-principles calculations to obtain a complete description of the electronic states resulting from the interaction between concentric chemically modified nanotubes, followed by electronic transport calculations to characterize the impact of B and N doping on micrometer long metallic DWNTs.

Self-consistent calculations of the electronic structure of doped DWNTs are performed with the localized orbital basis set of the {\tt SIESTA} density functional-based  code \cite{PhysRevB.53.R10441,0953-8984-14-11-302}. A tight-binding-like Hamiltonian, which presents the compactness required to study transport properties within the Green's function formalism can be obtained. To compute the Green's function of long disordered DWNTs, we first calculate the Hamiltonian of short doped nanotubes sections. By randomly assembling pristine and doped sections, long DWNTs with rotational and longitudinal disorder can be built.
To avoid any artificial backscattering effect in the region where the segments match, the supercell in the {\it ab initio} calculation is chosen long enough to avoid any overlap between the perturbed potentials caused by the doping atoms. By means of real-space renormalization techniques \cite{doi:10.1021/nl802798q,ISI:000276179400007}, the transport properties of micrometer long DWNTs can be evaluated. 

A double-$\zeta$ basis set within the local density approximation (LDA) approach for the exchange-correlation functional was used. Although van der Waals functionals are known to accurately describe the interaction between the concentric layers of pristine double-walled carbon nanotubes, LDA was considered a suitable functional since the dominating effects here are the new states coupling the concentric tubes upon chemical modifications. Also within the LDA the required level of accuracy in the description of large unit cells of $\sim$ 1000 atoms can be obtained at an affordable computational cost, which is $\sim$ 10 times less expensive than van der Waals functionals \cite{PhysRevLett.103.096102}. Atomic positions were relaxed with a force tolerance of 0.02 eV/\AA. The integration over the Brillouin zone was performed using a Monkhorst sampling of 1x1x4 k-points. The radial extension of the orbitals had a finite range with a kinetic energy cutoff of 50 meV. The numerical integrals were computed on a real space grid with an equivalent cutoff of 300 Ry.

The electronic transport calculations were based on the Landauer-B\"uttiker (LB) formulation of the conductance, which is particularly appropriate to study the probability of an electron to be transmitted or backscattered as crossing a one-dimensional device channel. The system is phase-coherent and all the scattering events occur in the channel, which is connected to two semi-infinite leads with reflectionless contacts. Within the LB scheme, one computes the transmission coefficients $T_n(E)$ for a given channel $n$, which gives the probability of an electron to be transmitted at a certain energy $E$ when it quantum mechanically interferes with the nanotube scattering centers. For a pristine CNT, $T_n(E)$ assumes integer values corresponding to the total number of open propagating modes at the energy $E$. The conductance is thus expressed as $G(E)={G_0}\sum_{n}T_n(E)$, where $G_0$ is the quantum of conductance. 

In the following, two armchair concentric CNTs will be noted as (n)@(m), where n stands for the index of the smaller tube. Any two armchair CNTs following the rule m=n+5 have an inter-wall spacing of 3.3 \AA, close to that of the van der Waals-bonded planar graphite sheets. Both tubes are assumed to be rigid cylinders with parallel axes. The minimal doped system was constructed by repeating a concentric armchair DWNT unit cell 13 times (3.25 nm) along the z-axis such that the geometric and energetic perturbations caused by a doping element vanish at the edges of the supercell. The entire systems of $\sim$ 1000 atoms were fully relaxed. Graphite-like B and N bonding configuration were considered so that each doping atom is bonded to 3 C atoms at first neighboring positions. Figure \ref{fig1}-a represents a (7)@(12) DWNT with two doping atoms in substitution of two C atoms at both coaxial sidewalls. Figure \ref{fig1}-b represents a (5)@(12) DWNT with a doping atom only at the outer sidewall. The distance between the concentric shells is of $\sim$ 5.5 \AA. The substitution of a C atom for either a B or a N atom entails a small geometrical distortion of the hexagonal network in the surrounding on the doping site, with slight increasing B-C and decreasing N-C bond lengths with respect to pristine C-C bond length in the pristine DWNT.  

\begin{figure*}[htp]
 \centering
 \includegraphics[width=0.55 \textwidth]{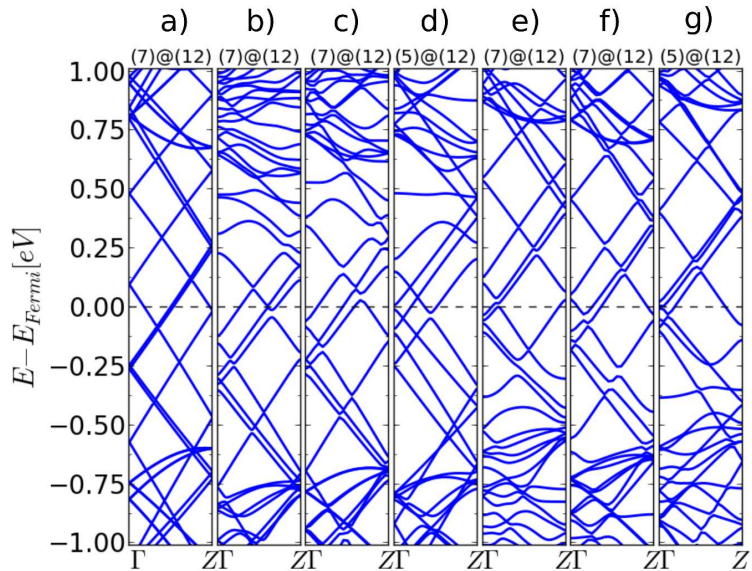}
 \caption{ a) Electronic band diagrams of double-walled carbon nanotubes with various types of doping. a) pristine 13-unit cell, b) one N atom at each concentric sidewall, and c) one N at the outer sidewall, of a (7)@(12) DWNT. d) one N at the outer sidewall, of a (5)@(12) DWNT. e) One B atom at each concentric sidewall, and f) one B at the outer sidewall, of a (7)@(12) DWNT. g) one B at the outer sidewall, of a (5)@(12) DWNT.   }
 \label{fig2}
\end{figure*}

\begin{figure*}[htp]
 \centering
 \includegraphics[width=0.95 \textwidth]{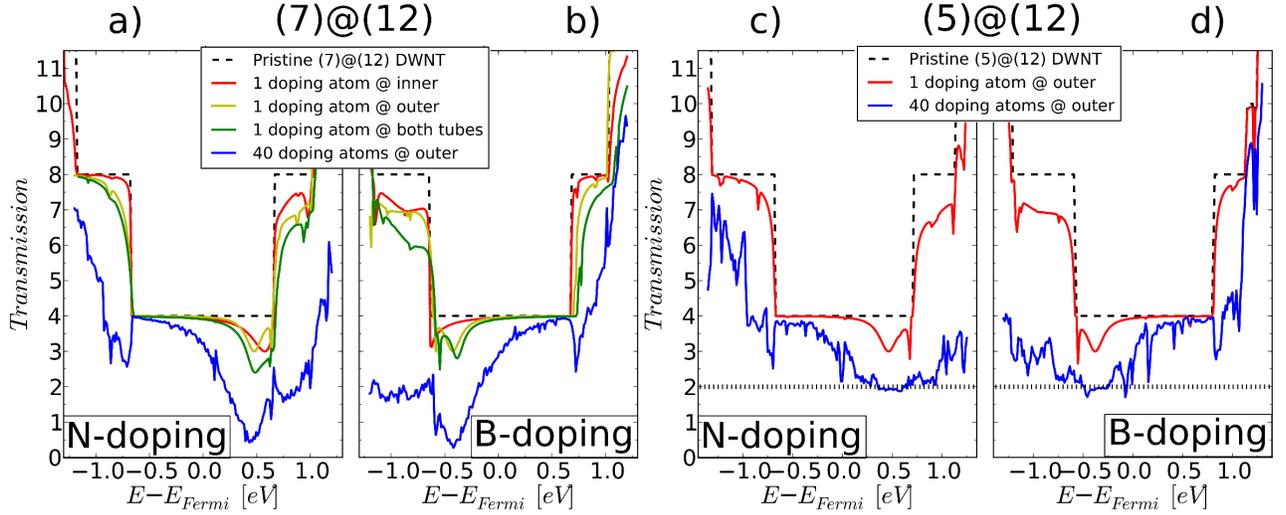}
 \caption{ a) Transmission coefficients for a (7)@(12) DWNT with substitutional N-doping in the inner or outer tube, and in both. As result of the quasi-bound states introduced on both tubes upon doping with 40 N atoms simultaneouly both coaxial tubes (as in Figure \ref{fig1}-a) a transmission drop that supresses the DWNT conductance ability at some energies is observed. b) same as in a) but for B-doping. c) Transmission coefficients for a (5)@(12) DWNT with substitutional N-doping in the outer tube uniquely. The transmission drop induced by 40 N-doping atoms barely decreases below T = 2 at some energies. d) same as in c) but for B-doping. The transmission coefficients for a pristine DWNT are in dashed lines. In c) and d), horizontal dotted lines indicate T(E) =  2  }
 \label{fig3}
\end{figure*}

A careful analysis of the electronic states of both pristine and doped DWNTs is crucial to understand the loss of conductance ability due to the chemical modification of the inner and outer shells. Self-consistent {\it ab initio} calculations were used to obtain accurate descriptions of the charge density distribution and the electronic structure of each system ground state.
Figure \ref{fig2}-a shows the electronic band diagram of a pristine 13-unit cell coaxial (7)@(12) DWNT. As widely reported, the system is metallic with several bands crossing the Fermi level originating from the Bloch sum of the $p_z$ carbon orbitals. For an isolated armchair single-walled CNT, $T_n(E)=2$ near the Fermi level (first plateau) and 4 for the second plateaus. Given that tube-tube interaction is weak, the conductance of a pristine DWNT in the ballistic conduction regime is given as the sum of all conducting channels of each tube at a given energy. The dashed lines in the conductance profile panels in Figure \ref{fig3} represent the transmission coefficients of a pristine DWNT, evolving from T = 4 close to the charge neutrality point (first plateau) to T = 8 in the second plateaus.

\begin{figure*}[htp]
 \centering
 \includegraphics[width=0.95 \textwidth]{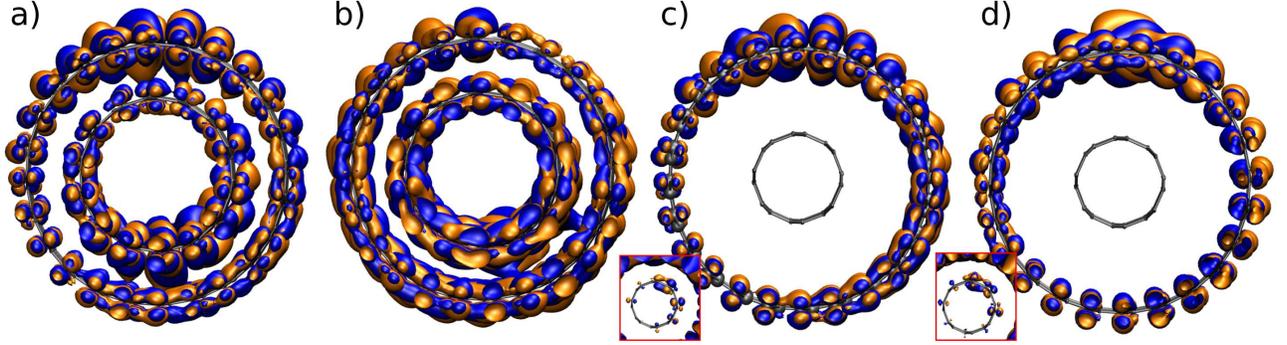}
  \caption{Electronic wave function plots at the $\Gamma$-point of the quasi-bound states at which the antiresonances are found in Figure \ref{fig3}. For a (7)@(12) DWNT one doping atom on both inner and outer tubes create a localized state in a) the electron band for N atoms, and b) in the hole band for B atoms. On the contrary, for a (5)@(12) doping on the outer tube does not affect the inner tube for both c) N atom, and d) B atom. Wave functions in main panels are plotted at an isosurface value of $1.6\times 10^{-2}$ e/\AA $^3$, and in insets of $4.1\times 10^{-4}$ e/\AA $^3$
}
 \label{fig4}
\end{figure*}

A N atom has an extra electron with respect to a C atom and behaves as a donor element, yielding an excess of charge density in the N-rich areas. Consequently n-type doping is expected to introduce electronic localized states above the Fermi level. Indeed, this is observed in the band diagram (Figure \ref{fig2}-b) of two coaxial (7)@(12) DWNT each with a N atom in substitution. Electronic states below the Fermi level remain practically unaltered whereas at higher energies a disruption of the $\pi$-conjugated network is observed with the apparition of small gaps and bands with narrow energy spread. The calculated conductance is presented in Figure \ref{fig3}-a. A transmission decrease, similar to the transmission profile reported for N-doped single-walled CNTs \cite{PhysRevLett.84.2917,PhysRevB.81.193411}, is observed in the electron band with a pronounced dip in conductance in the first plateau. Notice that, while the resonant backscattering of quasibound states in a single nanotube reduces the transmission in a quantum of conductance, for a two-fold N-doped DWNT the transmission drop is not the addition of the independent contributions of each tubes separately, i.e. T(E) $>$ 2 at the minimum. This suggests a mixing of the electronic states of the coaxial tubes that results on the creation of a bound state distributed on both tubes which quenches less than two transmission channels. This can be verified by plotting, for two substitutional N atoms in the positions shown in Figure \ref{fig1}-a, the wave function at the $\Gamma$-point of the dispersiveless electronic state at $\sim$ 0.5 eV above the Fermi level, (see Figure \ref{fig2}-b). Figure \ref{fig4}-a shows that the wave function spreads on both coaxial tubes, demonstrating that both the N-doping and the short inter-wall distance contribute to the mixing of both tubes electronic states, which in turn yields a reduction of the backscattering strength. It is worth to point out that the transmission profile of a N atom in substitution at either the outer tube or the inner tube is similar to that of an isolated single-walled CNT \cite{PhysRevLett.84.2917,PhysRevB.81.193411}, with a conductance drop equivalent to a transmission channel (see Figure \ref{fig3}-a). 

\begin{figure*}[htp] 
 \centering
 \includegraphics[width=0.95 \textwidth]{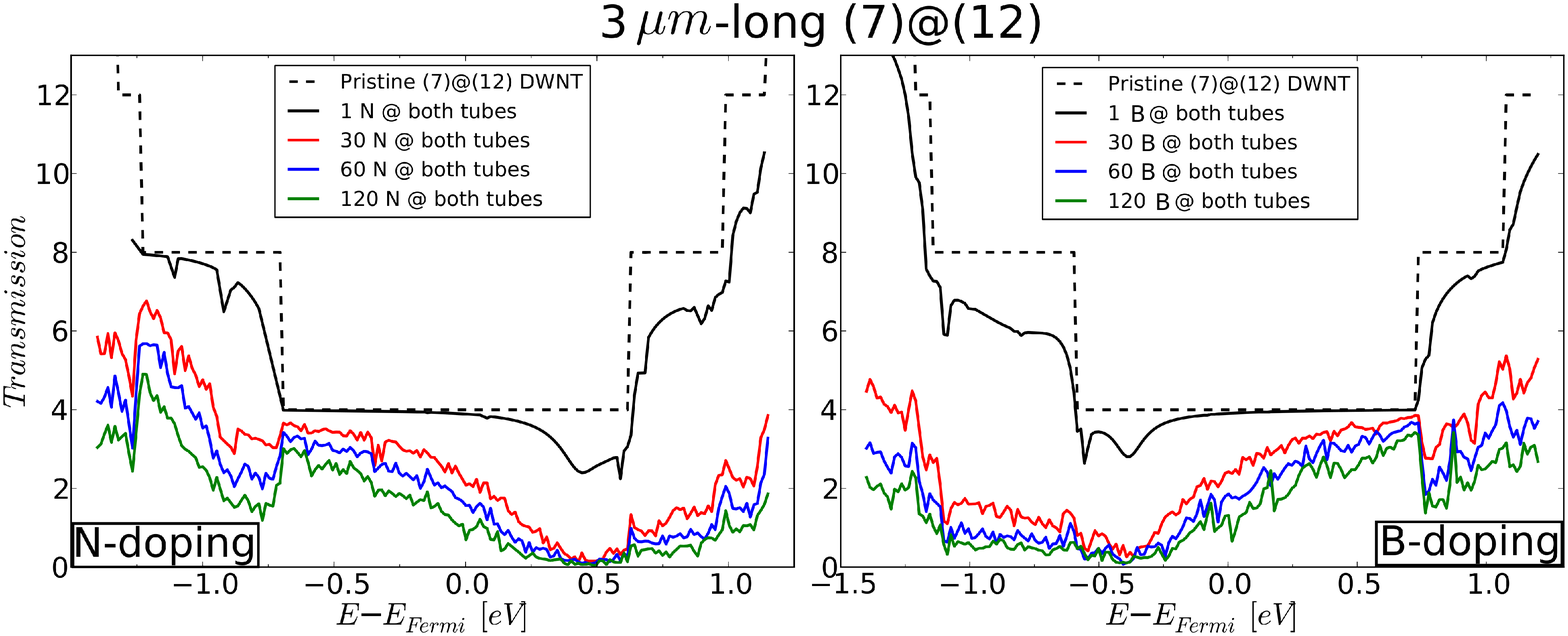}
 \caption{a) Averaged transmission coefficients for 3 $\mu$m-long (7)@(12) DWNTs with random distribution of substitutional N atoms on both coaxial tubes. The transmission for two N atoms, one on each tube, is shown with continuous black line. b) Same as in a) but for B-doping. The transmission coefficients for a pristine DWNT are in dashed lines. Curves have been averaged over 20 different random configurations. }
 \label{fig5}
\end{figure*}

One might assume then that a single doping atom is not enough to induce the hybridization of both coaxial tube states. Actually the band diagram of Figure \ref{fig2}-c shows that at $\sim$ 0.5 eV there is one electronic bands less than in the former case and some bands partially recover the dispersion. However the averaged transmission over 10 random configurations of 40 N-doped and pristine DWNT sections show that the coupling between both tube electronic states exists and enhances the backscattering efficiency, yielding a full suppression of the four available conduction channel at the resonance energy. This indicates that, similarly to functionalized DWNTs \cite{doi:10.1021/jp402355x}, inter-shell hybridization plays an important role in decreasing the conducting ability of the coaxial doped tubes, and that the wall-to-wall distance is a deciding factor for efficient control of electric current in chemically modified DWNTs. As a matter of fact, doping uniquely the outer tube and increasing the inter-wall separation leads to a practically complete restoration of the hyperconjugated carbon $\pi$-network of the inner shell, as Figure \ref{fig2}-d evidences with gapless and dispersive electronic bands. The flat band at the same energy at which Figure \ref{fig3}-c shows two conductance dips indicates the apparition of a quasi-bound state. The saturation of the transmission drop at T(E) = 2 of a (5)@(12) DWNT with 40 two-fold N-doped sections distributed in a random fashion suggests that the large inter-shell distance prevents the inner tube from being affected by the outer-sidewall modification. This is consistent with experimental results on DWNTs where covalent functionalization occurs strictly at the surface of the outer wall and the inner wall is essentially unaltered by the chemical modification \cite{doi:10.1021/nn201024u}. 
The plots in Figure \ref{fig4}-c and inset of wave function at the $\Gamma$-point of the flat band demonstrate that the electronic state spreads mainly over the outer tube. Only for very low values of the isosurface a small contribution of the inner tube to the state can be distinguished. This is related to the very small transmission drop observed below the T = 2 limit, pinpointing to a slight charge transfer between sidewalls as reported for a physisorption event \cite{doi:10.1021/nl0514386}. The small contribution of the inner sidewall to the total transmission decrease can be directly ascribed to a conventional diffusive process rather than to the presence of an electronic localized state on its surface. Notice that although the introduction of a doping atom in the carbon network keep the $sp^2$ character of the original hexagonal lattice and no $sp^3$ rehybridization is induced, the consequence on the inter-shell interaction and the hybridization of both tube states is greater than in $sp^3$ functionalized DWNTs \cite{doi:10.1021/jp402355x}.

Similar reasoning can be conducted to analyze the impact of B-doping on the electronic and transport properties of DWNTs. Owing to the acceptor character of a B atom embedded in a carbon network a deplexion of the charge density befalls in the surrounding of the doping site. An analysis based on the Mulliken population confirms that the B atoms at each of a (7)@(12) DWNT sidewalls gains a charge equivalent to $\sim$ 0.5 electrons at the expense of some charge of surrounding C atoms. The band diagram of Figure \ref{fig2}-e evidences that boron p-type doping distorts the bonding $\pi$-states of the coaxial tubes with several neighboring band mixing, whereas the $\pi^*$-states remain unaltered. This is concurrent with the transmission drops observed in the valence band and that, with respect to the N-doped DWNT, exhibit mirror symmetry around the charge neutrality point (see Figure \ref{fig3}-b). The wave function at the $\Gamma$-point of the electronic state at $\sim$ 0.37 eV below the Fermi level is plotted in Figure \ref{fig4}-b, and reveals that both tubes contribute to the total charge density at that energy, providing a strong hybridization character to this particular state. It is noticeable the marked resemblance with the transmission profile of a single B atom at the outer tube, and that the two dips in conductance are less pronounced than in the N-doped counterpart. Notwithstanding, succesive backscattering phenomena in a long DWNT with random distribution of several two-fold B-doped sections lead to a full suppression of the four available conduction channels at the resonance energy. As in the N-doping case, increasing the wall-to-wall distance the inner tube remains virtually unaffected by B-doping of the outer tube, as the band diagram (Figure \ref{fig2}-g), transport coefficients (Figure \ref{fig3}-d), and $\Gamma$-point wave function plots (Figure \ref{fig4}-d) evince.  

Current experimental techniques for the fabrication of B and N doped DWNTs do not allow for the selective incorporation of doping atoms in only one of the sidewalls \cite{ISI:000274004700001}. To further explore the electronic transport properties of doped metallic DWNTs, we have performed a mesoscopic study for realistic tubes with lengths of 3 $\mu$m and random distribution of an increasing number of dopants in both sidewalls. In Figure \ref{fig5} the quantum transmission coefficients of metallic (7)@(12) DWNTs with different concentrations of doping atoms are presented. Strong backscattering is observed at the resonance energies of the donor (acceptor) N (B) atom quasibound state as discussed above. The initially small conductance drop of a single impurity is significantly amplified when adding a larger number of doping atoms following the initial single atom signature. The accumulation of doping atoms along the DWNT enhances disorder effects and allows quantum interferences between scattering centers to set in. The conductance of the N-doped DWNT enters the localized regime in the conduction band at energies around 0.50 eV, and the transmission is fully suppressed (Figure \ref{fig3} left-panel). The transmission in the valence band is also deeply affected by the presence of the quasi-bound states but keeps a good conduction ability within the diffusive regime. Similarly, for B-doping a transmission drop is observed in the valence band (Figure \ref{fig3} right-panel), whereas the conduction band is less affected. Comparing the transmission profiles of both types of chemical doping, both the shape and the location of the transmission dips exhibit mirror symmetry with respect to the Fermi level. The multiple backscattering phenomena between the impurity atoms inevitably leads to a full suppression of the electronic transport capability of the coaxial tubes for a wide range of energies, suggesting the possibility of utilizing B and N atoms for the creation of transport gaps in multi-walled CNTs. 

In conclusion, it has been observed that the inter-wall interaction between coaxial nanotubes is greatly enhanced upon substitutional doping as a result of the hybridization of their electronic states. By means of a multi-scale approach based on first-principles and electronic transport calculations, transport regimes in doped DWNTs have been explored. In constrast to post-synthesis treatments of pristine DWNTs with chemical groups that only modify the outer wall and leave the inner wall unaffected, B and N doping represents an advantage in creating mobility gaps in metallic multi-walled CNTs. Both B and N atoms in substitution of C atoms at both sidewalls of the concentric tubes are an efficient source of disorder with the ability to suppress DWNT conductance at the valence (B atoms) and conduction bands (N atoms). Substitutional doping dramatically affects the performance of DWNTs and may be potentially used to control electron transmission in multi-walled based transistors.

This research used resources of the Argonne Leadership Computing Facility at Argonne National Laboratory,
which is supported by the Office of Science of the U.S. DOE under contract DE-AC02-06CH11357. 
This research also used resources of the National Center for Computational Sciences at Oak Ridge National Laboratory, which is supported by the Office of Science of the U.S. Department of Energy under Contract No. DE-AC05-00OR22725.

\end{document}